\documentclass[12pt,a4paper]{article} 

\usepackage{subfigure}
\usepackage{amsmath,amssymb,amsthm,mathrsfs,amsfonts,dsfont}
\usepackage{dcolumn}
\usepackage{bm}
\usepackage{braket}
\usepackage{physics}
\usepackage{mathtools}
\usepackage{diagbox}
\usepackage[utf8]{inputenc}
\usepackage[english]{babel}
\usepackage{scalerel}[2014/03/10]
\usepackage{stackengine}
\usepackage[noend,noline,ruled,linesnumbered]{algorithm2e}
\usepackage[noend]{algpseudocode} 
\usepackage{footnotebackref}
\usepackage{lipsum}
\usepackage{hyperref}
\usepackage{multirow}
\usepackage{makecell}
\usepackage{float}
\restylefloat{table}
\usepackage{placeins}

\hypersetup{colorlinks=true, linkcolor=blue, citecolor=blue, urlcolor=black }
\makeatletter

\usepackage{graphicx}
\usepackage{tikz}
\usepackage{qcircuit}
\usepackage[backend=bibtex]{biblatex}
\begin{document}

\title{Monte-Carlo Option Pricing in Quantum Parallel}


\author{
Robert Scriba\\
\textit{The University of Western Australia, Perth, WA 6009, Australia}
\and
Yuying Li\\
\textit{Cheriton School of Computer Science, University of Waterloo, Waterloo, Canada}
\and
Jingbo B. Wang\thanks{jingbo.wang@uwa.edu.au}\\
\textit{The University of Western Australia, Perth, WA 6009, Australia}
}
\maketitle

\begin{abstract}
    

\vspace{-0.2cm}
Financial derivative pricing is a significant challenge in finance, involving the valuation of instruments like options based on underlying assets. While some cases have simple solutions, many require complex classical computational methods like Monte Carlo simulations and numerical techniques. However, as derivative complexities increase, these methods face limitations in computational power. Cases involving Non-Vanilla Basket pricing, American Options, and derivative portfolio risk analysis need extensive computations in higher-dimensional spaces, posing challenges for classical computers.

Quantum computing presents a promising avenue by harnessing quantum superposition and entanglement, allowing the handling of high-dimensional spaces effectively. 
In this paper, we introduce a self-contained and all-encompassing quantum algorithm that operates without reliance on oracles or presumptions. More specifically, we develop an effective stochastic method for simulating exponentially many potential asset paths in quantum parallel, leading to a highly accurate final distribution of stock prices. Furthermore, we demonstrate how this algorithm can be extended to price more complex options and analyze risk within derivative portfolios.

\end{abstract}

\pagenumbering{roman}
\setcounter{page}{1}

\pagenumbering{arabic}
\setcounter{page}{1}

\vspace{-0.3cm}

\section{Introduction}
\vspace{-0.5cm}
Financial derivative pricing problems play a crucial role in the financial industry. These problems involve determining the fair value of financial derivatives, such as options, which derive their value from an underlying asset or financial instrument. In some special cases (simple European options) these problems can be solved analytically. Elsewhere classical computational methods, such as Monte Carlo simulations and numerical techniques, are commonly used to solve these pricing problems. However, as the complexity and desired pricing precision of financial derivatives increases, these methods face limitations in terms of computational power and efficiency. Some cases, such as pricing Non-Vanilla Basket, American Options and analysing risk of derivative portfolios involve higher-dimensional spaces, solving optimal stopping problems or require nested simulations. Classical computers struggle to efficiently handle the vast number of calculations required to model these complex systems accurately. \\

In these cases, quantum computing algorithms present a solution. Due to their exponential speedup and ability to handle high-dimensional spaces, quantum algorithms can provide faster and more efficient solutions for derivative pricing leading to more accurate pricing. This facilitates more confident pricing and more informed trading and hedging for hedge funds, market makers and other financial institutions. This can lead to tighter and more liquid markets and thus more trading, bringing with it larger financial incentives for these firms.

\vspace{-0.4cm}
\section{Quantum Finance and Option Pricing}
\vspace{-0.5cm}
With the rapid growth in quantum computing algorithm development, there has been significant research in the area of Quantum Finance. Research in quantum optimisation, stochastic modelling, quantum adiabatic computation, and quantum machine learning have all found applications to problems in the field of computational finance \cite{TangYehui2022Rpap, GomezAndres2022ASoQ}. For example, significant work has been done in the application of quantum optimisation to address the challenge in portfolio management. Promising approaches include adiabatic quantum computation \cite{GrantErica2021BQAC} as well as Quantum approximate optimisation algorithms (QAOA) and Quantum Walk based portfolio optimisation, as outlined by Slate et al \cite{SlateN2021Qwpo}.  
This paper extends our earlier work on quantum portfolio optimisation to quantum option pricing. In the following, we will first provide a review of prior work in this area.  

\vspace{-0.5cm}
\subsection{Quantum option pricing via QAE}
\vspace{-0.5cm}
Current quantum approaches to path-independent option pricing are mostly centred around the use of the Quantum Amplitude Estimation (QAE) algorithm~\cite{RebentrostPatrick2018QcfM, StamatopoulosNikitas2020Opuq, HerbertSteven2022QMCI, UdvarnokiZoltan2023QaoM}. These approaches encode the probability distribution and payoff function in the amplitude of an ancillary qubit register, where this amplitude is representative of the expected option payoff. They then utilise the aforementioned quantum algorithms in order to achieve a polynomial (quadratic) quantum speed up when compared to classical Monte-Carlo sampling. 

The first step of such an algorithm requires the loading of the final distribution of stock prices, {$S_i^T$}, where time $T$ the 'final' time refers to the time of expiry of the option. This loading of a final distribution is not trivial for many problems and is generally assumed in the literature to be achievable via some algorithm $A$, where 
\vspace{-0.3cm}
\begin{equation}\label{A_op}
   \mathcal{A}\ket{0}_{n} = \sum_{i=0}^{2^n - 1} {\sqrt{p_i}\ket{i}_n}.
\end{equation}
The addition of a rotation onto an ancillary qubit \cite{RebentrostPatrick2018QcfM} is used to apply the payoff function $f(S_T)$, denoted $f_i$, which in equation \ref{Af_op} is now included in A.
\begin{equation}\label{Af_op}
   \mathcal{A}\ket{0}_{n} \ket{0} = \sum_{i=0}^{2^n - 1} {\sqrt{1-f_i}\sqrt{p_i}\ket{i}_n}\ket{0} + {\sqrt{f_i}\sqrt{p_i}\ket{i}_n}\ket{1}.
\end{equation}

It is thus clear that the probability of measuring the ancillary qubit in the $\ket{1}$ state (\ref{P(1)}) is approximately equivalent to the expectation value of the option payoff (\ref{E(Op)}) with sufficiently large $2^n -1$. Note that to produce the option value the expected payoff is simply discounted by the risk free rate $r$ over time till expiry of the option $T$:
\vspace{-0.3cm}
\begin{equation}\label{P(1)}
    P(Ancilla = \ket{1}) = \sum_{i=0}^{2^n - 1} {f_i * p_i},
\end{equation}
\vspace{-0.2cm}
where
\vspace{-0.3cm}
\begin{equation}\label{E(Op)}
    \mathbb{E}(f(S_T)) = \int ^{\infty }_{-\infty }f\left( S_{T}\right) \cdot P\left( S_{T}\right) dS_{T}.
\end{equation}

When viewing the encoded state given in Eq.~\ref{Af_op}, it is clear that a repeated measurement of the ancillary qubit, analogous to sampling of a Bernoulli distribution \cite{UdvarnokiZoltan2023QaoM}, 
and counting the proportion of measurements giving the $\ket{1}$ will tend to the normalised expectation value. 
However, simply taking measurements of this state will not improve on the number of classical samples required to give the same accuracy. The Quantum Amplitude Estimation algorithm will need to be applied at this stage to achieve a quadratic speed-up over its classical counterparts~\cite{CarreraVazquezAlmudena2021ESPf}. 
The resulting sampling error of this method scales as $\mathcal{O}(1/N)$ \cite{LiYongming2023QMCa} where $N$ represents the number of applications of $A$, analogous to 'Quantum Samples'. Comparatively classical Monte-Carlo sampling results in error scaling of $\mathcal{O}(1/\sqrt{N})$ \cite{GlassermanPaul2002PVwH}.

As discussed above, current quantum algorithms for option pricing often operate under the assumption that the final probability distribution of stock prices is known and available in a functional form \cite{UdvarnokiZoltan2023QaoM}. This assumption raises significant challenges, particularly when dealing with real-world financial models.
Requirements for loading such a distribution onto a quantum circuit are that the functional form of the distribution is known and is efficiently integrable \cite{GroverLov2002Cstc} or alternatively is prepared via a quantum circuit. Notably, the former requirement is met in cases like the Black-Scholes-Merton model, where analytical forms are readily accessible. It is however crucial to consider cases where analytical forms of the stock distribution are not known, as in known cases the option value of a simple option can be directly analytically solved removing any need for Quantum or numerical approaches or equivalently if a classical process is used to determine an approximation for the distribution. As such, the true complexity arises when financial models lack analytical forms or when they aim to capture intricate market dynamics that extend beyond the BSM assumptions.

Rebentrost et al. \cite{RebentrostPatrick2018QcfM} provided valuable insights by demonstrating the possibility of uploading log-concave distributions, such as the Gaussian distribution of $W_T$, onto quantum circuits. This achievement aligns with scenarios involving single stochastic variable processes for individual stocks. However, it is imperative to recognise that the broader landscape of financial derivatives pricing encompasses a multitude of stock price distributions that often lack analytical forms.
Efforts by Stamatopoulos et al. \cite{StamatopoulosNikitas2020Opuq} underscore the pressing need to efficiently represent distributions of financial parameters on quantum computers, especially when explicit analytical representations are absent. Their critical observation that loading arbitrary states into quantum systems typically demands exponentially many gates \cite{PleschMartin2011Qpwu} highlights the impracticality of modeling arbitrary distributions as quantum gates and demonstrates that developing a quantum circuit to appropriately reflect stock price probability distributions is not trivial.

These collective insights accentuate the formidable challenge associated with constructing quantum circuits that faithfully reflect stock price probability distributions, particularly when these distributions lack simple analytical forms. Stamatopoulos et al. \cite{StamatopoulosNikitas2020Opuq} propose a potential solution in the form of a quantum Adversarial Networks (qGANs) approach, which works by leveraging known data samples to fit a representative distribution using a polynomial number of gates. It amplifies the need for further exploration and innovation in this domain. It is important to note that this method still, as with simpler cases, requires some known information about the final stock price distribution. In the case of qGANs the method required data points of the stock price distribution to be produced, which may require some classical or quantum Monte-Carlo simulation in and of itself. 
Vasquez and Woerner \cite{CarreraVazquezAlmudena2021ESPf} provided one method of state preparation and highlight the requirment for such an algorithm. This area of research is however underrepresented and other approaches, which are more computationally efficient in the classical domain, have not been investigated.

We also note that there are significant similar problems in finance, which in large have not been addressed by research into Quantum Monte Carlo approaches. Ons such example is nested risk valuation. Initial research into risk valuation with quantum algorithms, such as by Woerner and Egger \cite{WoernerStefan2019Qra}, take a single simulation approach which may not extend well to risk valuation for times other than expiry. Classically, nested Monte-Carlo simulations have been adopted in industry in valuing guarantees for VA portfolios, yet efficient nested simulation for large varied portfolios has proven difficult due to prohibitively expensive computation\cite{GanGuojun2015Volv}. Quantum approaches to this problem may hold an efficient solution to nested simulation risk valuation problems.

In summary, the gap in the current research landscape resides in the efficient representation of complex, non-analytical financial asset value distributions on quantum computers and the development of full quantum algorithms which do not rely on the loading of known distributions. As well as the extension of such algorithms to risk valuation and other more computationally intensive financial pricing problems. This underrepresented challenge is central to the accurate estimation of option contract values and risk assessment in real-world financial scenarios. The development of quantum algorithms and methods capable of effectively handling such distributions remains under-researched area of literature and an open avenue for future research.


\vspace{-0.5cm}
\subsection{Monte-Carlo Option Pricing in Quantum Parallel}
\vspace{-0.5cm}
We propose a method to address the issue of loading probability distributions which are not efficiently integrable, a problem under-addressed in the literature and a crucial step assumed possible by many previous works. The proposed algorithm simulates stock price paths as in a classical Monte-Carlo simulation, encoding stock price and variable values in the computational basis and simulating paths in Quantum Parallel. 

It is helpful to consider the classical Monte-Carlo Simulation approach for European Option pricing using a Black-Scholes Stochastic evolution. In this approach time to expiry of the option is discretised into $m=T/\Delta t$ time steps. Each Monte-Carlo simulation begins with the initial stock price $S_0$. At each time step for each path the stock price evolves stochastically according to
  $S_t$ = $S_{t-1}$  + drift term + diffusion term, where the drift term = $\mu \times \Delta t \times S_{t-1}$, $\mu$ is the drift rate, and 
  the diffusion term = $\sigma \times \Delta W_{t-1} \times S_{t-1}$, $\sigma$ is the volatility of the stock, and $\Delta W$ is a stochastic variable, which is normally distributed about mean = 0 and std.= $\sqrt{\Delta t}$. Noting that each stochastic variable is independent and each path or Monte-Carlo simulation is independent of all others. A payoff function is then applied to all final stock prices. For a call option this follows a simple piece-wise function
  \vspace{-0.2cm}
  \begin{equation}\label{callpayoff} 
  f\left( S_{T} \right) =
    \begin{cases}
    S_{T} - k & \text{if } S_{T} > k \\
    0 & \text{if } S_{T} \leq k
    \end{cases}
  \end{equation} 
 where $K$ represents the strike price of the option. This simulation is run N times where thre mean payoff of all paths is taken to give the calculated option price.


Before discussing the Analogous Monte-Carlo in Quantum Parallel Circuit it is useful to consider an alternative presentation of the classical Monte-Carlo simulation of a single variable Itô's process evolution for option pricing. In this method no loops are explicitly used and instead a matrix is populated with variables such that a row product can be taken, giving the same result as a single stock price evolution (one path):
\begin{equation} \label{stochevolve}
\begin{split}
    S_{t} &= S_{t-1} + \mu \Delta t \cdot S_{t-1} + \sigma \Delta W \cdot S_{t-1} \\
          &= S_{t-1} \cdot (1 + \mu \Delta t + \sigma \Delta W) \\
    S_{T} &= S_{0} \cdot \prod_{i=1}^{m} \left( 1 + \mu \cdot \Delta t + \sigma \Delta W_i \right)
\end{split}
\end{equation}
    

In order to replicate the desired Monte Carlo simulation in quantum parallel all steps of the process must be efficiently carried out with unitary quantum gates. Below we consider the approach of producing each of the $N$ path's $\Delta W_{t}^i$ values, for some $t$, in parallel. Followed by calculating their respective $S_{t}^i$ values in parallel and then repeating both these operations for each time step until time of expiry ($T$) is reached, at which point the payoffs for each path $P^i$ are calculated in parallel. We then propose encoding the mean in the amplitude of an anicllary qubit, as in previously outlined QAMC methods. From here QAMC methods may be used to ensure quantum speed up, demonstrating the proposed MCQP method as a solution to state preparation problems in QAMC solutions for Option Pricing. The algorithm approach is laid out in algorithm \ref{MCQP}.

\begin{algorithm}[htbp]\label{MCQP}
\caption{Quantum Monte Carlo Simulation}
\KwResult{Mean Option Payoff}
\For{$t$ in $1$ to $T/\Delta t$}{
        Generate $\Delta W^i$ random bit strings with operator $\mathcal{R}$\;
        Calculate $S^i$ bit-string values from $\Delta W^i$ and previous $S^i$ with operator $\mathcal{T}$\;
}
Calculate $P^i$ bit-string values from final $S^i$ with operator $\mathcal{P}$\;
Encode Square root of mean of $P^i$'s in amplitude of ancillary qubit with operator $\mathcal{M}$\;
Measure Amplitude, utilise QAMC amplitude estimation for speed up\;
\KwRet{Mean Payoff Value}
\end{algorithm}

\begin{figure}[htp]
    \centering
    \includegraphics[width=10cm]{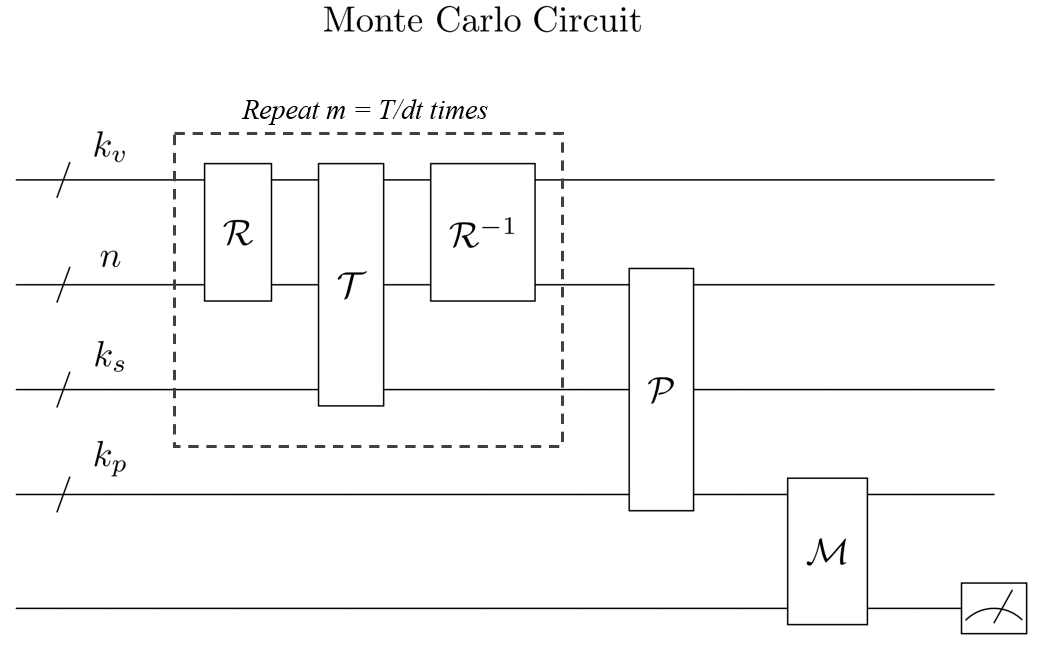}
    \caption{Quantum Circuit for Monte-Carlo simulation in Quantum Parallel}
    \label{fig:circuit1}
\end{figure}

Represented in Figure \ref{fig:circuit1} is the proposed general quantum circuit structure for the execution of MCQP for the case of a single stochastic variable (following GBM) previously outlined. The circuit requires four quantum registers and one ancillary qubit. 

Three of the registers (the Stochastic Variable Register, the Stock Price Register and the Payoff Register) are denoted to be of size $k_v$, $k_s$ and $k_p$ qubits respectively. The choice of k corresponds to the precision of the respective numeric values encoded within the computational basis for each register. For example, for a given $k_s$ qubits, the superposition of stock prices for some time step $t$ is encoded in a length $k_s$ bit-register, effectively discretising or 'binning' stock prices into $2^{k_s}$ values which extends to the Stochastic Variable and Payoff Registers as in equation \ref{discretise} where $DF(x)$ represents the discritised numeric value of $x$,
\begin{equation}\label{discretise}
     DF(\Delta W^s_t) = \ket{\Delta W^s_t}_{k_v}, \ DF(S_t) = \ket{S_t}_{k_s}, \  DF(P) = \ket{P}_{k_p}
\end{equation}

The other register, the Index Register, is of size $n = log(N)$ qubits, where $N$ represents the number of index values, ensuring there are $N$ independent arithmetic process each simulating a potential stock price and payoff application, outlined in following sections of the paper. Thus $N$ can be considered the effective number of paths simulated in the Monte-Carlo. 


The final ancillary qubit is a single qubit initialised in a $\ket{0}$ state. This qubit's amplitude encodes the mean payoff of the option and hence can be measured to determine the final Option Price. Amplitude estimation techniques outlined by previous works on QAMC can then be used to achieve quadratic speed up.



In order to effectively simulate independent paths of a MCMC Black Scholes evolution a randomly sampled $\Delta W^i$ stochastic variable for each path for each time must be produced. In other words each $\Delta W_{t}^i$ bit-string value produced must be sampled from the appropriate normal distribution and must be independent of any other $\Delta W_{t}^i$. In order to achieve this with the application of $\mathcal{R}$ we propose generating a seed value used for the sampling of a uniform distribution. The seed generation function involves utilising a logistic map of r=4 to create approximate chaos and is shown in algorithm \ref{seedgen}, where $i$ is the index value and $m$ is the time step. The requirement to produce different samples for each time step requires that for a unitary operation an additional register of $log(m)$ qubits is required as input for $R$ this register is excluded from here on for conciseness.

\begin{algorithm}\label{seedgen}
\caption{Seed Generation Algorithm (labelled 'seed')}
\KwData{Positive integers $i$, $m$; Non-negative integer $k$}
\KwResult{Seed value as an integer}
$x_0 \leftarrow \frac{1}{\sqrt{i}^3}$\;
$x \leftarrow x_0$\;
\For{$\text{iter} \gets 1$ \KwTo $k$}{
    $x \leftarrow 4 \cdot x \cdot (1 - x)$\;
}
\textbf{Let} $y \leftarrow \left(\left(x \cdot 10000\right) \mod 10\right) \cdot 1000000 + m \cdot 100$\;
\KwRet{$y$}\;
\end{algorithm} 

Using the random seed, a number is sampled from a uniform distribution with a value range of reasonable upper and lower bounds of $\Delta W$. If the value sampled from the uniform distribution is accepted according to the method outlined below, the value is accepted as the $\Delta W^i_t$ value. However, if it is rejected a new seed is regenerated by increasing k which continues until a sample value is accepted. This function was simulated for 1000 $i$ values and 100 $m$ values according to the recursive algorithm \ref{seedar}.

\begin{algorithm}\label{seedar}
\caption{Random Sample Generation with Rejection Sampling (labelled 'seedar')}
\KwData{Initial values $i$, $m$, and $k$; Maximum number of retries $max\_retries$}
\KwResult{Accepted sample}
\If{$k \geq max\_retries$}{
    \textbf{Raise} Exception('Reached maximum number of retries ($max\_retries$).')
}
Generate a random number $p\_act$ uniformly distributed between 0 and 1\;
Generate a seed value using the function $seed(i, m, k)$\;
Set the random seed to $seed\_value$\;
Generate a random sample $sample$ uniformly distributed between -4.5 and 4.5\;
Calculate $norm\_value$ as $\frac{\text{PDF of } sample \text{ with } \mu=0 \text{ and } \sigma=1}{0.4}$\;
\If{$norm\_value \leq p\_act$}{
    $k \leftarrow k + 1$ \;
    \textbf{Return} $seedar(i, m, k)$ \;
}
\Else{
    \textbf{Return} $sample$ \;
}
\end{algorithm}

\begin{figure}[!ht]
    \centering
    \includegraphics[width=15cm]{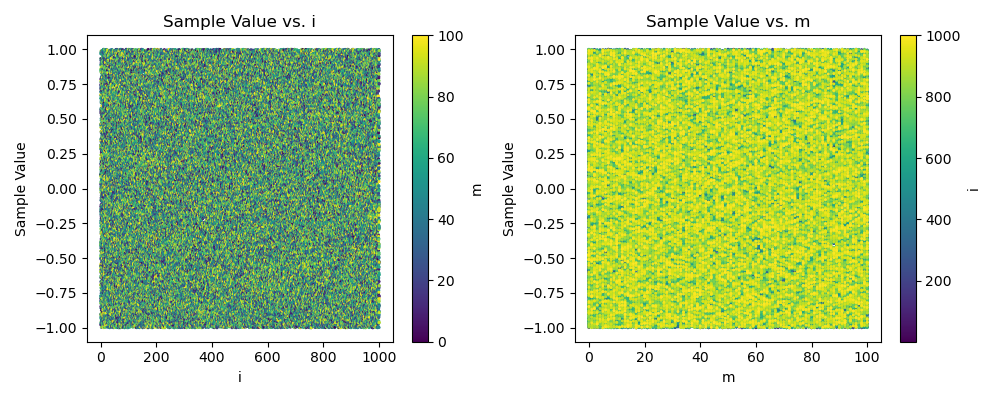}
    \caption{Random Sample via Seed Generation Algorithm}
    \label{fig:randseedunif}
\end{figure}

\begin{figure}[!ht]
    \centering
    \includegraphics[width=15cm]{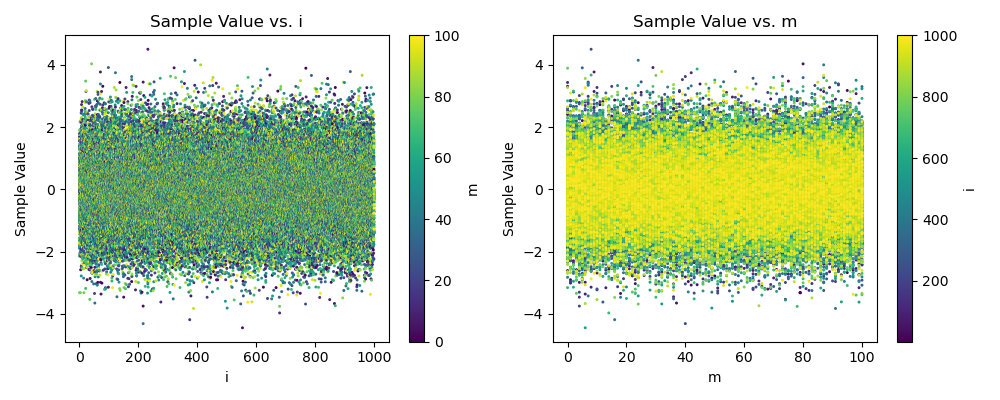}
    \caption{Random Sample Generation with rejection Sampling}
    \label{fig:randseednorm}
\end{figure}

The results of the simulations run with algorithm \ref{seedgen} and algorithm \ref{seedar} are shown in figure \ref{fig:randseedunif} and \ref{fig:randseednorm} respectively. Further testing on data should be conducted in order to more scrutinously determine if there exists some underlying structure within results. However, visually these result appear uniformly and normally random respectively.


Based off of a simple acceptance/rejection method, the superposition of values, in the register is effectively altered to reflect the desired distribution. In the case of our GBM for the BSM model the desired distribution for $\Delta W$, as previously stated, follows a normal distribution of mean = 0, std. = $\sqrt{\Delta t}$. As shown in equation \ref{stochevolve}, for the BSM case, the value of $1+\mu \times \Delta t +\sigma \times \Delta W$ can alternatively be produced as the 'Variable' value according to a normal distribution of mean = $1 + \mu \times \Delta t$, std. = $\sigma \times \sqrt{\Delta t}$. This choice of variable may simplify the $\mathcal{T}$ operation, but the former provides a more intuitive extension to more complex cases and is assumed for further steps.\\

\begin{figure}[!ht]
    \centering
    \includegraphics[width=15cm]{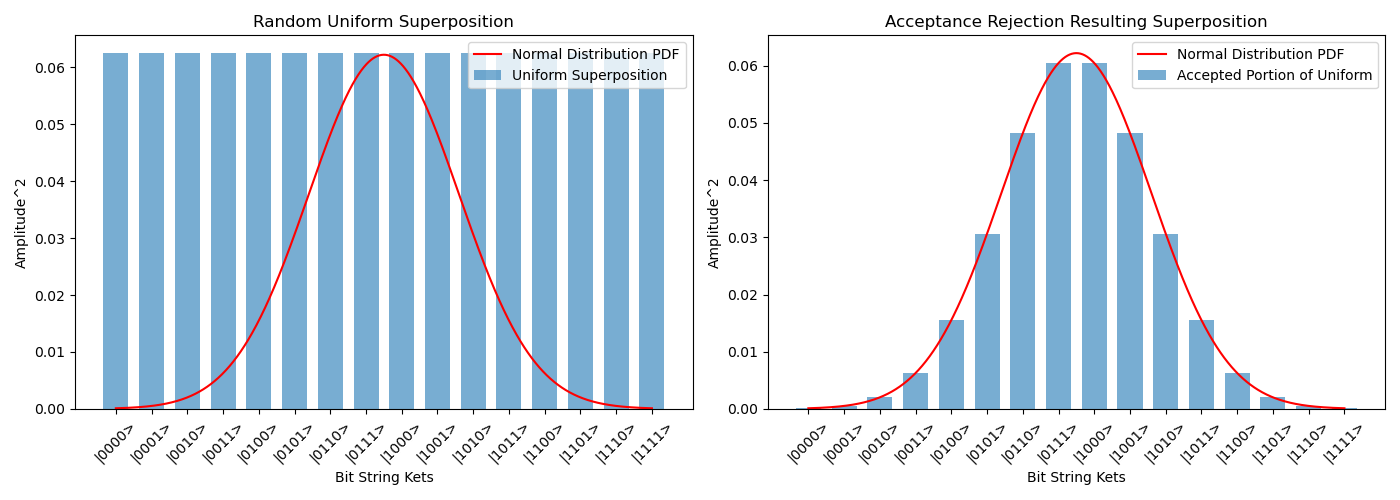}
    \caption{Acceptance Rejection Diagram: 4-Qubit Example}
    \label{fig:accrej}
\end{figure}

A graphical representation of the effective trimming of a uniform distribution to a normal distribution is shown in figure \ref{fig:accrej}. In which the sampled value is accepted with probability equal to the desired distribution's normalised density value and rejected with probability of one less the acceptance probability. For the case of the standard normal distribution $\mu = 0, \ \sigma = 1$ the density value is divided by $0.4$ to normalise.


The application of the random number generation, as stated, is denoted by the application of an operator $\mathcal{R}$ which acts on a state as follows:
\begin{align}
\mathcal{R}: \ket{i}_n\ket{0}_{k_v} &\mapsto \ket{i}_n\ket{\Delta W_{t}^i}_{k_v}
\end{align}
Thus, for some state $\ket{\phi_{t}} = \sum_{i=0}^{N} \ket{i}_n\ket{0}_{k_v}\ket{S_{t}^i}_{k_s}$ we have:
\begin{align*}
\mathcal{R}: \ket{\phi_{t}} &= \sum_{i=0}^{N}\ket{i}_n\ket{\Delta W_{t}^i}_{k_v}\ket{S_{t}^i}_{k_s} = \ket{\psi_{t}}.
\end{align*}
Hence, the inverse operator $\mathcal{R}^{-1}$ will act as follows,
$    \mathcal{R}^{-1}: \ket{i}_n\ket{\Delta W_{t}^i}_{k_v} \mapsto \ket{i}_n\ket{0}_{k_v}.$

The Time Step Evolution step is carried out via the application of a quantum operator $\mathcal{T}$. In this step the stock price of each path $\ket{S_t^i}$ at some time $t$ is evolved to $\ket{S_{t+1}^i}$, depending on the relevant stochastic variable $\ket{\Delta W_t^i}$ according to the BSM,
\begin{equation}
    S_{t+1}^i= S_{t}^i \times (1 + \mu \Delta t + \sigma \Delta W_{t}^i).
\end{equation}
\vspace{-0.3cm}

As $S_{t+1}^i$ is dependent on $S_{t}^i$ and $\Delta W_t^i$, an index register $\ket{i}_n$ is used to entangle $\ket{S^i}$ and $\ket{\Delta W^i}$ states only, meaning that each  of the $N$ paths evolve independently in parallel. The application of the time step evolution operator $\mathcal{T}$ is outlined as,
\begin{equation}
     \mathcal{T} : \ket{i}_n\ket{\Delta W_{t}^i}_{k_v}\ket{S_{t}^i}_{k_s} \mapsto \ket{i}_n\ket{\Delta W_{t}^i}_{k_v}\ket{S_{t+1}^i}_{k_s}.
\end{equation}

The application of the random number generator function $\mathcal{R}$ followed by the time step operator $\mathcal{T}$ act on the $\ket{\phi_t}$ state as follows and then the inverse $\mathcal{R}^{-1}$,  i.e.   $\mathcal{R}^{-1}\mathcal{T}\mathcal{R}: \ket{\phi_{t}} \mapsto \ket{\phi_{t+1}}.$ 
The application of $\mathcal{R}^{-1}\mathcal{T}\mathcal{R}$ is repeated $m = T/\Delta t$ times, evolving $S_0$ to the superposition of $\ket{S_T^i}$ stock price states.


Once the stock price register has been evolved to its final expiry time — that is, each path has reached its end $\ket{S_T^i}$ — the payoff function must be applied to each stock price. This payoff function for the European call option follows the piecewise equation~\ref{callpayoff}, meaning
\begin{equation}
P^i = f\left( S_{T}^i \right) =
\begin{cases}
S_{T}^i - k & \text{if } S_{T}^i > k \\
0 & \text{if } S_{T}^i \leq k
\end{cases}.
\label{callpayoff}
\end{equation}
Equivalently, $P^i = \max\left(S_{T}^i - k, 0\right)$ and
the payoff $P^i$ is calculated from $S_{T}^i$ as:
\begin{equation}
\mathcal{P} : \ket{\phi_{T}}\ket{0}_{k_p}  \mapsto \sum_{i=0}^{N} \ket{i}_n\ket{0}_{k_v}\ket{S_{T}^i}_{k_s} \ket{P^i}_{k_p}
\mapsto \sum_{i=0}^{N}\ket{\phi_{T}^i} \ket{P^i}_{k_p}.
\end{equation}

The operator $\mathcal{M}$ encodes the normalised square root of the mean payoff into the amplitude of the $\ket{1}$ state of the ancillary qubit. It acts on the relevant state as follows,
\begin{equation}
    \mathcal{M} :  \sum_{i=0}^{N}\ket{\phi_{T}^i}\ket{P^i}_{k_p}\ket{0} \mapsto   
\sum_{i=0}^{N}\ket{\phi_{T}^i}\ket{P^i}_{k_p} \left( \sqrt{\frac{1-P^i}{N}}\ket{0} + \sqrt{\frac{P^i}{N}} \ket{1}\right).
\end{equation}
The full quantum algorithm can now be written as follows,
\begin{equation}  
        \mathcal{M}\mathcal{P}(\mathcal{R}_{-1}\mathcal{T}\mathcal{R})^m: \ket{\phi_0}\ket{0}_{k_p}\ket{0} \mapsto
\sum_{i=0}^{N}\ket{\phi_{T}^i}\ket{P^i}_{k_p} \left( \sqrt{\frac{1-P^i}{N}}\ket{0} + \sqrt{\frac{P^i}{N}} \ket{1}\right).\\
\end{equation}

Note that after the application of $\mathcal{M}$ quantum amplitude estimation techniques outlined by papers discussing an QAMC algorithm approach can be utilised to achieve polynomial quantum speed up. Due to the fact that previous literature on QAMC has been centred around amplitude estimation, there may be potential to create a more efficient overall algorithm by considering alternative procedures than encoding the mean payoff and utilising QAE to measure the amplitude. 

It is known that for classical Monte Carlo simulations, error scales as $\mathcal{O}\left(1/\sqrt{N}\right)$, where $N$ is the number of simulations. 
Previous results have highlighted the ability of QAMC, given efficient loading of final stock price distributions, to outperform classical methods with error scaling as $\mathcal{O}\left(1/M\right)$ where $M$ is the number of applications of operator applications, analogous to 'Quantum Samples'. The proposed MCQP algorithm ensures this speed up can be achieved by providing means for efficient state preparation. For MCQP the simulated number of paths $N$ requires $n$ index qubits and requires only $m$ repetitions of $\mathcal{R}_{-1}\mathcal{T}\mathcal{R}$ where $m$ represents the number of time-steps. The required number of time steps and precision qubits is generally independent of the number of dimensions and simulations and provides only a linear overhead in complexity. The size of the quantum circuit is effectively $\approx$ $n + k_p + (k_v + k_s ) \cdot d + log(m) + 1$ and the depth is $\mathcal{O}(m)$.\\

Preparing arbitrary quantum states has exponential complexity with respect to the number of qubits \cite{CarreraVazquezAlmudena2021ESPf}. Comparatively our method has linear complexity with respect to the number of qubits. We find that the addition of $d$ dimensions, in other words, stochastic variables, requires the addition of $2 \cdot d$ quantum registers of size $k_d$ and $k_{v_{d}}$ which represent the number of precision qubits for the additional dimension and its stochastic variable respectively.

\vspace{-0.3cm}
\section{Results and Discussion}
\subsection{Monte-Carlo Performance versus Simulation Parameters}

Initial simulations were conducted for two separate models accross a range of option strike prices ($k$ values) and asset volatility ($\sigma$) values and compared to a BSM analytically solved theoretical call option price. Fixed simulation parameters were initial stock price = 100.0, drift rate = 0.05, total time = 1.0, number of time steps = 100, number of price bin bits=5, number of simulations= 10000. \\

\begin{figure}[htbp]
    \centering
    \includegraphics[width=17cm]{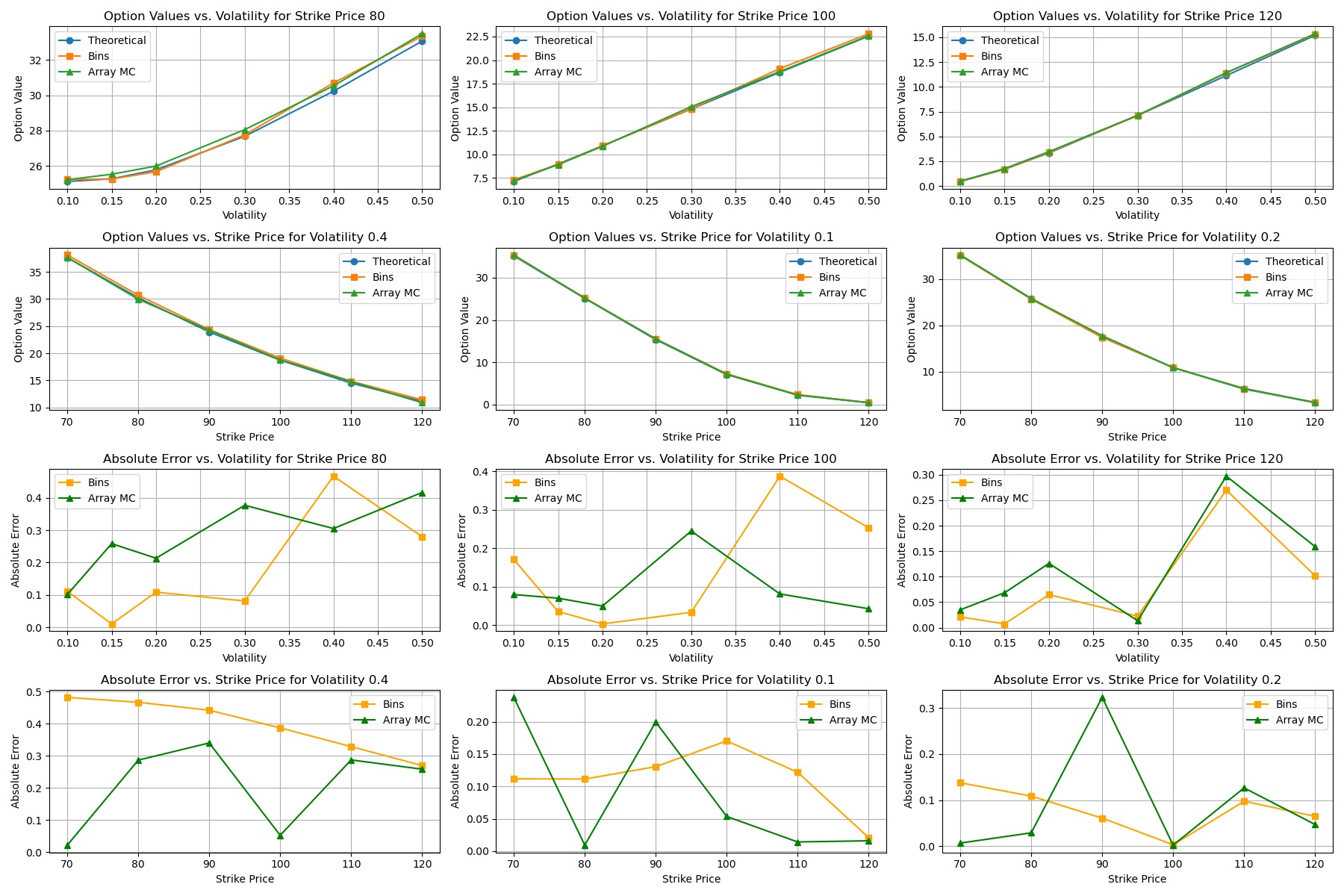}
    \caption{Option Value for Strikes, volatility}
    \label{fig:initialsim}
\end{figure}

The first model, denoted 'Bins' in figure \ref{fig:initialsim}, is carried out by taking a theoretical final stock price distribution and discretising stock price values into $2^n$ bins, where $n$ is the number of price bin bits. The midpoint value of each bin is then used as the approximate stock price for that bin, after which the payoff function is applied and the average is calculated. This process emulates the discretisation required in the aforementioned QAMC pricing methods \cite{UdvarnokiZoltan2023QaoM}, for cases where the final stock price distribution can be efficiently loaded. The process was carried out with only $2^5$ stock price bins. The second process, denoted 'Array MC' in figure \ref{fig:initialsim} represents the number of effective paths set to $1e4$ approximately equivalent to $13$ index bits.

Figure \ref{fig:initialsim} shows that pricing of both models can be considered similarly accurate. The number of approximate price bin bits ($5$) is significantly lower than the number of path index bits ($13$) despite producing similar absolute errors. This implies, that when a final stock price distribution can be efficiently loaded, the computational cost required for storing discretised stock price values is initially not significant. More significantly, when varying strike price, there are no significant trends in absolute pricing error. This is inline with expectations as the stock price simulation and binning of prices itself is independent of strike price. A general upward trend in absolute error in pricing with an increase in volatility values is observed. In the case of Array MC, this error may increase due to the increased contribution of simulation values further from the initial stock price having greater theoretical impact on pricing. i.e. due to increased volatility final stock prices are more distributed and the space becomes more sparse. For the Bins method similar reasoning may result in worsening pricing accuracy as bin midpoint prices become less representative of the impact of wider payoff value bins. 

\subsection{Algorithm Pricing Accuracy: Circuit Depth and Size}
\begin{figure}[htp]
    \centering
    \includegraphics[width=17cm]{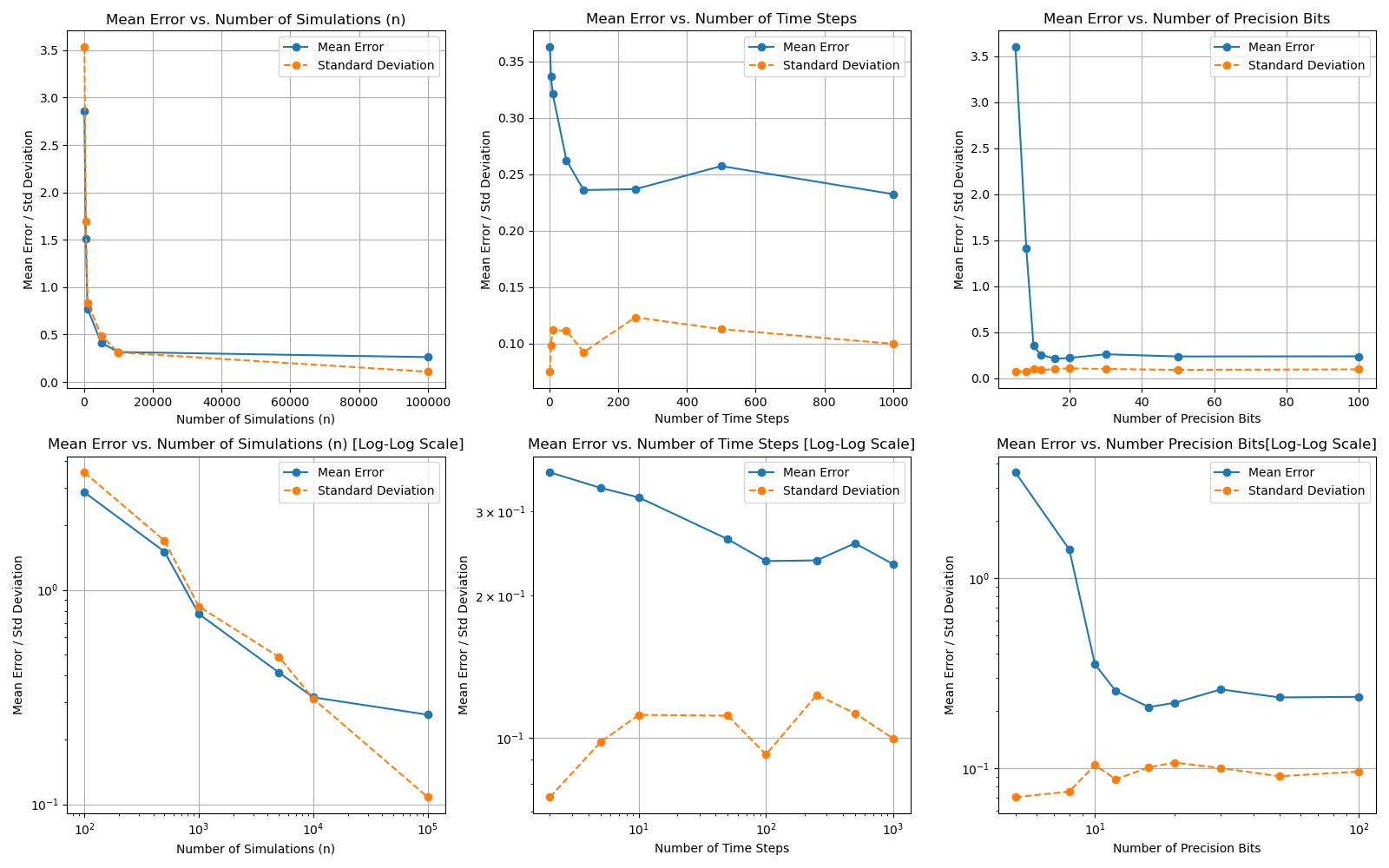}
    \caption{Error to theoretical value vs number of simulations, time steps and bit precision}
    \label{fig:errorvar}
\end{figure}

The construction of the MCQP circuit requires a number of quantum registers including the Index Register, Variable Register, Stock Price Register and Payoff Register. The algorithm also requires the repeated application of the operators $\mathcal{R}\mathcal{T}\mathcal{R}^{-1}$ which are applied $m$ times. In order to get a better practical understanding of how the size of quantum registers and the number of time step repetitions may impact the overall accuracy of the MCQP method for pricing options, classical simulations were conducted with results displayed in \ref{fig:errorvar}. These simulations involved calculating an option price under the BSM, given a set of initial parameters, using a slightly modified classical Monte-Carlo simulation in which parameter values were 'rounded' in order to approximate the effect of encoding stochastic variable values with some number of qubits in the computational basis. In addition the product of parameter values was calculated via a for loop, where at each stage product values are also 'rounded' for the same purpose. Thus numeric values were approximately rounded to the nearest of $2^{precision bits}$ values, equally spaced in between an approximated upper and lower bound for the value. Consider a stock price with price bounds of 0 and 300, i.e. the probability of the stock price being outside bounds is $\approx$ 0. Then, given some number of $precision bits$ $k$, the numeric stock price value will be rounded to the nearest of the $2^k$ values between 0 and 300. For this simplified simulation, the same number of precision bits were used for parameter values, stock price values and payoff values. This modified algorithm was used to investigate the impact of changing the number of precision bits on the accuracy of the pricing. 

Simulation carried out with the following input parameters: initial-stock-price = 100.0, drift-rate = 0.05, volatility = 0.4, total-time = 1.0, strike-price = 100, num-trials = 30\\

\begin{figure}[htp]
    \centering
    \includegraphics[width=16cm]{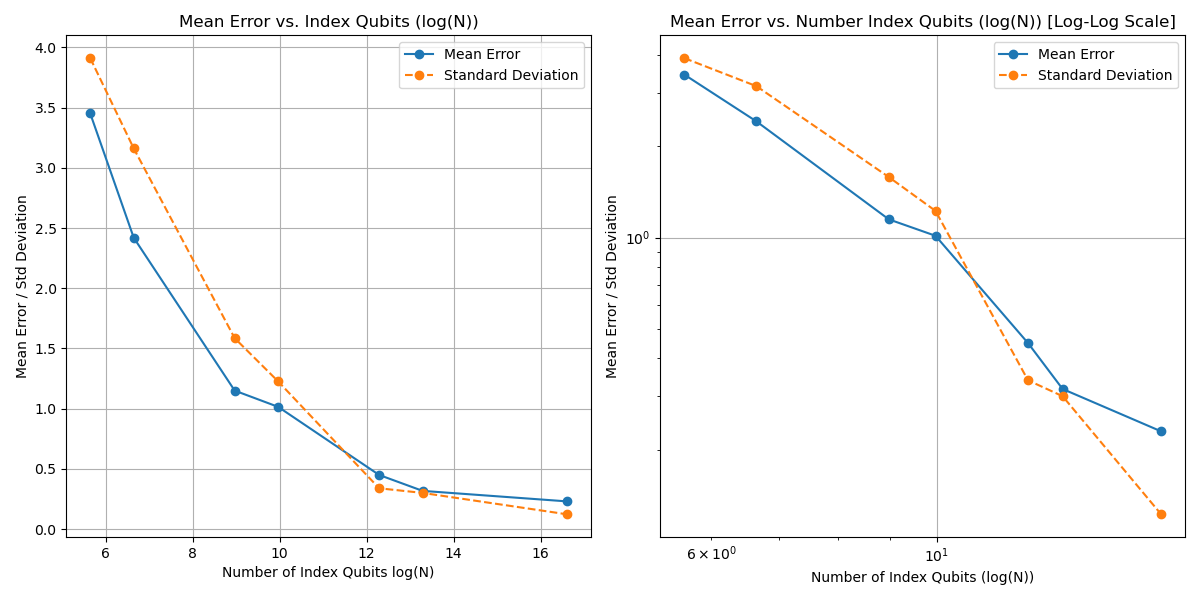}
    \caption{Error to theoretical value vs Index bits = $log(N)$}
    \label{fig:errorlogn}
\end{figure}

We can observe a strong correlation between the number of simulated paths ($N$) and pricing error, as well as pricing standard deviation. This correlation demonstrates an approximate gradient of $-1/2$ on the log-log plot, thus represents an approximate relationship of,
$    Error \propto 1/N^2. $ 
This is in agreement with the well known scaling of Monte-Carlo error. It should be noted that the mean error of trials for 10000 simulations does not adhere to the expected relationship. This may be the result of error from the number of time steps, set to the maximum tested number of time steps (1000), or error from the precision of Numpy stored values which is approximately equivalent setting precision bits to 32. The discrepancy between mean error and standard deviation implies the presence of systematic error which requires further investigation. It is important to consider that the number of simulations is not directly proportional to the number of qubits required, but rather the equivalent number of simulations of the MCQP algorithm is given by $N = 2^{n qubits}$ where $n qubits$ is the number of index qubits. Considering the behaviour of pricing error in relation to the log of the number of simulations shown in \ref{fig:errorlogn}, it can be seen that pricing error scales down significantly quicker, as would be theoretically expected. \\

\textbf{Accuracy vs. Number of Time Steps}\\
There is a clear correlation is present between mean pricing error and number of time steps. Interestingly, the impact of number time steps on the standard deviation of results is very minimal. , we expect error to scale with $delta t$ according to Euler's approximation. This also follows intuitively, as with a low number of time steps the final stock price function will be more normal in shape. Comparatively, increasing the number of time-steps ensures the final distribution tends toward log-normal as required for the BSM case. However, error begins to plateau at approximately 100 time steps. As the plateau is at a mean error value equivalent to that reached by all variable simulations, we are not able to comment on the cause or the mean error's relationship with the number of time steps within this plateau range. Further simulations, either classically or on quantum hardware are required to gain further insight into this relationship.\\

\textbf{Accuracy vs. Number of Precision Bits}\\
Viewing the plots of error vs. number of precision bits we can see a sharp initial improvement in result accuracy, viewing the data on the log/log scale plot we can see that the rate of decrease of mean error declines on both scales, as a result it may be possible that within the context of a MCQP process the relationship between the pricing error and number of precision bits is exponentially decaying. However, there is insufficient data to support this and further simulations should be done in order to fit the approximate relationship. In addition it should also be noted that the standard deviation was relatively constant and in fact increased, this may follow from the idea that very low precision numbers would result in more consistent final stock values.\\

Further simulations and research will likely bring more clarity to the effect of changing the number of time steps on the mean pricing error and standard deviation of results. If pricing performance is indeed shown to plateau with the number of time steps, then the depth of the quantum circuit required for MCQP can be kept practically shallow. In other words the number of repetitions $m$ can be kept relatively low. Further investigation should be conducted to extract an approximate effective relationship between pricing error and the number of precision bits for each register.

\section{Extension to useful/more complex cases}
The majority of use cases for solving financial derivative problems with quantum algorithms are problems with higher dimensionality that cannot be analytically solved. These problems can suffer from the curse of dimensionality and demonstrate areas where a quantum speed up may be significant.

In order to demonstrate the potential of a Monte-Carlo simulation in Quantum parallel as a solution to more complex and relevant problems, we consider a range of natural extensions corresponding to current prevalent higher dimension pricing problems.

\vspace{-0.5cm}
\subsection{Heston Model}
\vspace{-0.5cm}
The extension to evolving stock prices using the Heston model demonstrates extension to simulating models with multiple correlated stochastic variables. Model's with multiple stochastic variables are well know to be capable of more accurately representing true stock behaviour.

When considering a Monte-Carlo simulation following the more simplistic single stochastic variable as in the BSM model, the stock price at any time $t$ can be factorised out of the value of the stock for this path on the next time interval, shown by equations \ref{stochevolve}. This allows us to rewrite a final stock price of a path as a product of independent values. The addition of an another stochastic variable, one which is not independent of the first stochastic variable, such as in the Heston model may make such an equation impossible. This is the case in the Heston Model, in which the second stochastic variable, responsible for Brownian motion within volatility, alters the volatility value \ref{volevolve}. This ensures the stock price is no longer factorisable as a product of all time steps without first calculating the matrix of volatility values. The stochastic evolution of volatility for each time step itself is also not factorisable into a product of all entries. This means that a quantum approach must have 'communication' from the volatility register to the stock-price register. After volatility values are calculated, for a specific time step, as a superposition of values in the computational basis these values impact the volatility values of next time step (on the same register) as well as the values of the stock prices for the next time step (stock price register). 
Crucially, we note that regardless of the number of stochastic variables, correlations between them all paths remain independent, ensuring that exponentially many 'path simulations' can be conducted in Quantum parallel. It is thus expected that with the addition of a number of correlated stochastic variables, the required number of qubits grows linearly, with the addition of a $k$ qubit register for each variable where $k$ is responsible for the precision of the variable value. We therefore have
\begin{equation}
    S_t = S_{t-1} + \text{drift term} + \text{ diffusion term}
\end{equation} 
\vspace{-0.3cm}
\begin{equation} 
  \text{drift term} = \mu \times \Delta t \times S_{t-1} 
\end{equation} 
\vspace{-0.3cm}
\begin{equation} 
\text{diffusion term} = \sigma_{t-1} \times \Delta W_{t-1}^S \times S_{t-1}
\end{equation} 
where $\sigma_{t-1}$ is the volatility equal to the square root of variance $\sigma = \sqrt{\nu}$  that evolves according to \begin{equation}\label{volevolve}
  \nu_t = \nu_{t-1} + k (\theta - \nu_{t-1}) \times \Delta t + \xi \times \sqrt{\nu_{t-1}} \times \Delta W_{t-1}^{\nu}.
  \end{equation}
Here $\Delta W^S$ is the stochastic variable responsible for the Brownian motion of the stock price (as in the BSM model), $\Delta W^{\nu}$ is the stochastic variable responsible for the Brownian motion of the volatility. It should be noted that $W_t^S$ and $W_t^{\nu}$ have correlation $\rho$.\\


In order to extend the MCQP algorithm for the two variable Heston Model case, algorithm \ref{MCQP} can be altered by the addition of two steps for each time step. The first is the generation of volatility stochastic variable bit strings $\Delta W_v^i$ with correlation $\rho$ to $\Delta W^i$ and the second is the evolution of volatility values $\sigma^i$ which evolve according to equation \ref{volevolve}. In addition, the evolution of $S^i$ , in the case of the Heston Model, is not only dependent on the previous $S^i$ and the $\Delta W^i$ value but also the $\ket{\sigma^i}$ value. The adjusted algorithm in given below \ref{MCQP-H}.\\

\begin{algorithm}[htbp]\label{MCQP-H}
\caption{MCQP- Heston Model}
\KwResult{Mean Option Payoff}
\For{$t$ in $1$ to $T/\Delta t$}{
        Generate $\Delta W^i$ and $\Delta W_v^i$ random bit strings with operator $\mathcal{R_H}$\;
        Calculate $\sigma^i$ bit-string values from $\Delta W_v^i$ and previous $\sigma^i$ with operator $\mathcal{V}$\;
        Calculate $S^i$ bit-string values from $\Delta W^i$, previous $S^i$ and $\sigma^i$ with operator $\mathcal{T_H}$\;
}
Calculate $P^i$ bit-string values from final $S^i$ with operator $\mathcal{P}$\;
Encode Square root of mean of $P^i$'s in amplitude of ancillary qubit with operator $\mathcal{M}$\;
Measure Amplitude, utilise QAMC amplitude estimation for speed up\;
\KwRet{Mean Payoff Value}
\end{algorithm}

\begin{figure}[htp]
    \centering
    \includegraphics[width=12cm]{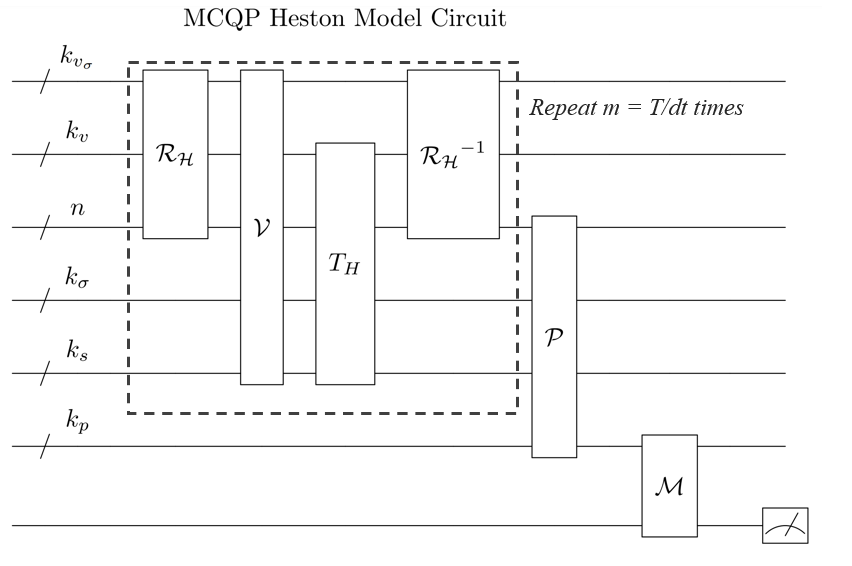}
    \caption{Quantum Circuit for Heston Model MCQP}
    \label{fig:hestoncircuit}
\end{figure}

Thus, the extension to the MCQP general circuit requires the addition of two quantum registers; the Volatility Stochastic Variable register, which is of size $k_{v_{\sigma}}$ qubits and the Volatility register, which is of size $k_{\sigma}$ qubits where, once again. The size of
these registers determines the precision of bounded values represented by the bit-strings of the register. It is also important to note that the operation of $\mathcal{R_H}$ and $\mathcal{T_H}$ is slightly different to the previous $\mathcal{R}$ and $\mathcal{T}$. Consequently the notation has been changed. One more addition to the circuit is the $\mathcal{V}$ gate, responsible for the evolution of volatility values. The full extended circuit can be seen in figure \ref{fig:hestoncircuit}. Importantly, the impact of an additional stochastic variable on the quantum process does not impact the number of time step repetitions and appears to result in a linear increase in overhead in creating the final distribution. Demonstrating the potential of MCQP processes for higher dimension problems.

The extension to the bi-variate case requires only linear increase in circuit size only. We note the full process acts as,
\begin{equation}
         \mathcal{M}\mathcal{P}(\mathcal{R_H}^{-1}\mathcal{T_H}\mathcal{V}\mathcal{R_H})^m: \ket{\phi_0}\ket{0}_{k_p}\ket{0} \mapsto 
\sum_{i=0}^{N}\ket{\phi_{T}^i}\ket{P^i}_{k_p} \left( \sqrt{\frac{1-P^i}{N}}\ket{0} + \sqrt{\frac{P^i}{N}} \ket{1}\right)\\
\end{equation}
Allowing for the same amplitude measurement methods to be used as with the single stochastic variable (BSM) case.
 
\subsection{Single split risk analysis}
One other significant application of Monte-Carlo simulation in Quantum finance is in risk evaluation. As previously stated, classical solutions to this problem grow quickly in computational complexity. The extension of a Monte-Carlo in quantum parallel requires, similarly to the classical form, a nested Monte-Carlo simulation.\\ 

We consider at first a single nested Monte-Carlo simulation for one option as it is outlined in \cite{GanGuojun2015Volv} and \cite{BroadieMark2011EREv}. Here the first 'simulation' stochastically evolves $N$ paths until time $\tau$, where time $\tau$ represents the time at which the portfolio (in this case a single option) value is of concern to risk assessment. This simple case is applicable in the case an institution desires to determine the probability of some lower portfolio value bound being crossed at time $\tau$. \\

For each of the $N$ original paths another 'simulation' is conducted from $\tau$ until expiry in order to determine the portfolio value (value of the option), at each potential point at time $\tau$. Each of these secondary Monte-Carlo simulations simulates $m$ paths. At this point the proportion of $N$ paths which have a portfolio value, at time $\tau$, of less than the threshold, represents the desired probability, as all paths are equally probable. This method can be easily altered for different risk metrics, for example to calculate expected profit and loss quantiles or percentiles. Such metrics are crucial for institutions which must adhere to risk limits.

\begin{algorithm}[htbp]
\caption{Nested Monte-Carlo Simulation for Portfolio Value at Time $\tau$}
\KwData{Number of outer simulations $N$, Number of inner simulations $m$, Threshold value $\epsilon$, Stock start price $S_0$, Option specifications}
\KwResult{Estimated probability $P(\text{Portfolio Value} < \epsilon)$}

$P(\text{Threshold Crossed}) \leftarrow 0$ \;
\For{$i \leftarrow 1$ \KwTo $N$}{
    Simulate 1 stock price path starting at $S_0$ from time $0$ to $\tau$\;
    Simulate $m$ paths starting at $S_{\tau}$ from time $\tau$  to $expiry \ (T)$\;
    Apply payoff function to the $m$ $S_T$ values
    Calculate the portfolio value at time $\tau$ as the discounted mean of the $m$ paths \;
    \If{Portfolio Value $< \epsilon$}{
        Increment $P(\text{Threshold Crossed})$ \;
    }
}
Estimated probability $P(\text{Portfolio Value} < \epsilon) \leftarrow \frac{P(\text{Threshold Crossed})}{n}$ \;
\KwRet{Estimated probability $P(\text{Portfolio Value} < \epsilon)$}
\end{algorithm}

For practical cases of portfolios with many financial derivatives with different expiry's, over different underlying assets, classical nested Monte Carlo becomes quickly very computationally intensive. The required modelling of multiple correlated stocks represents a multidimensional stochastic problem, which for the case of option pricing, will increase the required number of samples. In addition, the application of various payoff functions at different times increases the computational resources required for pricing risk.

Let us first consider the simpler, but still useful to solve, problem of a non-nested Monte Carlo simulation for calculating some value at risk (VaR) metric. This estimation can be simply achieved through the application of different payoff function which applies the option payoff and the risk metric function to all $N$ stock values at time T. This is similar in approach to the VaR pricing method outlined by Woerner et al \cite{WoernerStefan2019Qra}. For the example where the desired metric is the probability of some final option payoff being less than some value $\epsilon$, the payoff function $f$ is replaced with the function $f_r$,


\begin{equation}
f_r\left( S_T^i \right) = r \left( f \left( S_T^i \right) \right) =
\begin{cases}
1 & \text{if } S_{T} > k + \epsilon \\
0 & \text{if } S_{T} < k + \epsilon
\end{cases}
\end{equation}

where

\begin{equation}
f\left( S_{T} \right) =
\begin{cases}
S_{T} - k & \text{if } S_{T} > k \\
0 & \text{if } S_{T} \leq k
\end{cases}
\end{equation}

and

\begin{equation}
r\left(f\left(S_{T}\right)\right) =
\begin{cases}
1 & \text{if } f\left(S_{T}\right) > \epsilon \\
0 & \text{if } f\left(S_{T}\right) < \epsilon
\end{cases}
\end{equation}

Following the same steps as the simple pricing case of MCQP, the expectation value of this metric can be estimated. Described as the proportion of paths (equivalent to the probability) of some option payoff $\epsilon$ being exceeded at expiry.

The extension of MCQP to calculate risk metrics at some time $\tau$ prior to expiry is somewhat more involved and can be achieved with a 'nested' Monte Carlo approach in quantum parallel. A preliminary scheme for approaching this problem is layed out below with further research recommendations made later in the paper. This extension requires the addition of a register $\ket{j}_n$ which is labelled the secondary index register. Each of the $N$ unique $\ket{j}_n$ bit strings correspond to one of the $N$ secondary paths. Consequently with the entangled index registers $\ket{i}_n\ket{j}_n$, $N^2$ equivalent paths are represented from the point in time $\tau$ on-wards. As for the option pricing application the payoff function is applied, giving the resulting state,
\begin{equation}
    \ket{\phi_T^{ij}}\ket{P^{ij}}_{k_p} = \ket{i}_n \ket{j}_n \ket{0}_{k_v} \ket{S_{T}^{ij}}_{k_s} \ket{P^{ij}}_{k_p}
\end{equation}

From here we will drop the less relevant variable and stock price registers $\ket{0}_{k_v} \ket{S_{T}^{ij}}_{k_s} $. To this state, we apply a similar mean encoding, quantum digital to analogue conversion operator, as for option pricing. This operator applied to the previous state with an ancillary qubit, results in the following state transformation,
\begin{equation}
    \mathcal{QDAC}1: \sum_{i=0}^{N}\sum_{j=0}^{N}\ket{i}_n \ket{j}_n \ket{P^{ij}}_{k_p}\ket{0} \mapsto 
\sum_{i=0}^{N}\sum_{j=0}^{N}\ket{i}_n \ket{j}_n \ket{P^{ij}}_{k_p} \left( \sqrt{\frac{1-P^{ij}}{N}}\ket{0}_i + \sqrt{\frac{P^{ij}}{N}} \ket{1}_j\right)
\end{equation}

A quantum analogue to digital transformation operator is then applied, dropping the $\ket{j}$ index and $\ket{P^{ij}}_{k_p}$ registers, resulting in the following state change where the sum over $j$ of the square root of the payoff over N for each $i$ is labelled as $(\mathbb{E}P)^i$,
\begin{equation}
\mathcal{QADC}:\sum_{i=0}^{N}\sum_{j=0}^{N}\ket{i}_n \ket{j}_n \ket{P^{ij}}_{k_p} \left( \sqrt{\frac{1-P^{ij}}{N}}\ket{0}_i + \sqrt{\frac{P^{ij}}{N}} \ket{1}_j\right)  
\mapsto \sum_{i=0}^{N}\ket{i}_n \ket{(a\mathbb{E}P)^i} .
\end{equation}
This is followed by the application of the risk measure function $RF$,
\begin{equation}
    \mathcal{RF}: \sum_{i=0}^{N}\ket{i}_n \ket{(a\mathbb{E}P)^i} \mapsto \sum_{i=0}^{N}\ket{i}_n \ket{(\mathbb{E}P)^i} \ket{RF\left((\mathbb{E}P)^i \right)}
\end{equation}

Finally, another quantum digital to analogue conversion operator is applied, encoding the risk measure values for each of the $i$ paths in an ancillary qubit, where
\begin{equation}
\begin{split}
    \mathcal{QDAC}1: & \sum_{i=0}^{N}\ket{i}_n \ket{(\mathbb{E}P)^i} \ket{RF\left((\mathbb{E}P)^i \right)} \mapsto \\
    & \sum_{i=0}^{N}\ket{i}_n \ket{(\mathbb{E}P)^i}  \ket{RF\left((a\mathbb{E}P)^i \right)} 
    \\
    & \left( \sqrt{\frac{1-(\mathbb{E}P)^i}{N}}\ket{0} + \sqrt{\frac{(\mathbb{E}P)^i}{N}} \ket{1}\right)
\end{split}
\end{equation}

From here, once again amplitude estimation methods can be used to achieve efficient sampling of the risk measure encoded in the amplitude of the $\ket{1}$ state of the ancillary qubit.

The generation of many random numbers in quantum parallel is a requirement for the MCQP algorithm. We have initially outlined a potentially novel approach to this problem in the implementation of the operator $\mathcal{R}$. Further work should be done to demonstrate the efficiency of this operator and its effectiveness at producing random numbers. Further research could also be conducted to investigate other potential use cases for a generalised seed based quantum random number generation algorithm.

\section{Conclusion}
In this paper, we have explored the domain of financial option pricing, delving into classical approaches and recent quantum Monte-Carlo research efforts. A central theme has been the efficient preparation of asset price distributions at option expiration. We analysed relationships between model inputs and required computational resources and found initial indications of lesser dependency on numerical precision and time steps compared with the number of simulations. We have introduced a novel stochastic method for simulating asset paths in quantum parallel, illustrating its potential applicability to complex option pricing and portfolio risk analysis. This method represents an approach which, in its classical counter part is significantly more efficient than the classical counter parts of quantum methods of state preparation, potentially leading to savings in computational resources for certain option pricing problems. Furthermore, we have identified avenues for complementary research, including efficient quantum state preparation methods~\cite{GreenWang2025} and expanding the scope of quantum Monte-Carlo techniques in financial applications~\cite{WuLiWang2024}, as well as attempting further optimisation of the MCQP method.


\printbibliography

\end{document}